\documentstyle[fl eqn,twoside]{article}
\def\be{\begin{equation}}
\def\ee{\end{equation}}
\def\bi{\bibitem}

\begin{document}
\title{Quantum Mechanical Probability Interpretation In The 
Mini-superspace Model Of Higher Order Gravity Theory}
\author{Abhik Kumar Sanyal}

\maketitle

\noindent

\begin{center}
Dept. of Physics, Jangipur College, Murshidabad,
\noindent
India - 742213\\
\noindent
and\\
\noindent
Relativity and Cosmology Research Centre\\
\noindent
Dept. of Physics, Jadavpur University\\
\noindent
Calcutta - 700032, India\\
\noindent
e-mail : aks@juphys.ernet.in\\
\end{center}

\noindent
\begin{center}
\bf{\Large{Abstract}}
\end{center}
It has been shown that introduction of higher order curvature invariant 
terms like $R^2$ or $R_{\mu\nu}R^{\mu\nu}$ in the Robertson-Walker 
minisuperspace model of the Einstein-Hilbert action leads to 
Schr$\ddot{o}$dinger-like equation, whose corresponding 
effective Hamiltonan is hermitian. Thus, it is possible to 
write the continuity equation in a straight forward 
manner which reveals a quantum mecchanical probability 
interpretation of the theory.

PACS .04.50+h,04.20.Ha,98.0.Hw
\section{\bf{Introduction}}
In a couple of recent communications \cite{a:p} and \cite{a:c}, it has 
been observed that if the Einstein-Hilbert action is replaced or modified 
by the introduction of a curvature squared term in the Robertson-Walker 
minisuperspace model, then one of the true degrees oh freedom is 
distangled from the kinetic part of the canonical variables.This kinetic 
term behaves as a time parameter in the quantum domain and the corresponding 
Wheeler-deWitt equation looks like Schr$\ddot{o}$dinger equation.This 
result is interesting since it tends to resolve the long standing 
disagreement between `General Theory of Relativity' and ` Quantum 
Mechanics'. The disagreement is, in GTR time stands on the same footing 
as the space coordinates and it does not act as an exterrnal parameter, 
while `Quantum Mechanics' knows to predict uniquely only when time is 
seperated from the rest of the variables and acts as an external parameter. 
To resolve the contradiction Wheeler \cite{j:r} proposed that dynamical 
objects are space and not space-time, and the geometrical configuration 
of space evolves with time. This proposal requires a (3+1) decomposition 
of the metric and one can canonize the theory to get Wheeler-deWitt 
equation. However it is a very special feature of gravitation that it 
does not naturally lead to a Hamiltonian, but only to the Hamiltonian 
constraint. So it remained obscure to construct a satisfactory Hilbert 
space, on which the probability interpretation is based. In the quantum 
domain of the modified (by the curvature squared term) cosmological model, 
one of the canonical variables is automatically selected as the time 
parameter, resolving the disagreement.
\par
This essentially is not a new result. Horowitz \cite{h:p} and Pollock 
\cite {p:n} observed it two decades back. what new is, a proper choice of 
the canonical auxilliary variable turns out the effective Hamiltonian 
hermitian and thus it becomes possible to construct the equation of 
continuity and hence to establish the quantum mechanical idea of the 
probability and the current densities at least in the periphery of the 
Robertson-Walker minisuperspace model. Further, it gives rise to an 
effective potential whose extremization yields classical constraint equation 
\cite{a:p} and \cite{a:c}. These results have not been observed earlier, 
since it requires as mentioned, the correct choice of the auxilliary 
variable. The question thus naturally arises is, why a definite choice 
of such variable is required and how can it be chosen?      
\par
Canonical quantization of a theory whose action contains higher order 
terms is far from being trivial. It is required to choose an auxilliary 
variable, and to write the action in the canonical form. In the context 
of quantum cosmology, introduction of $R^2$ term in the action, requires 
such technique. Auxilliary variables can be chosen in many different ways 
\cite{h:p}, \cite{s:p} and \cite{a:p}. For all such variables it is 
possible to express the action in the canonical form, leading to the 
same classical field equations. However quantum dynamics turns out to be 
different for 
different auxilliary variable. Though such an important fact has only 
been observed fairly recently, it is not surprising. The reason being 
that, though all canonically related variables lead to the same classical 
field equations, there is no counterpart of such result in the quantum 
domain and hence should not be extended naturally. Rather, as mentioned, 
one has to choose such auxilliary variable in a unique manner so that 
correct quantum dynamics emerges. Boulware et-al \cite{b:a} prescribed a 
technique to pick up canonical variables for higher order theory. The 
prescription can be summarised in the following three steps. First, take 
the first derivative of the action with respect to the highest derivative 
of the field variable present in the action and idetify it as the 
auxilliary variable. Second, write the action in the canonical form after 
introducing the new variable and finally, quantize with basic variables.    
\par
The above technique was taken up by Hawking and Luttrell \cite{h:n}, 
Horowitz \cite{h:p} and Pollock \cite{p:n} to identify the auxilliary 
variable. Interestingly enough, Pollock \cite{p:n} showed that if one 
introduces auxilliary variable in the induced theory of gravity, where it 
is not at all required, one still ends up with a Schr$\ddot{o}$dinger 
like equation. Sanyal and Modak \cite{a:p} have shown that auxilliary 
variable can even be introduced in the vacuum Einstein-Hilbert action, 
yielding completely wrong quantum dynamics. Auxilliary variable can be 
introduced in such models if one simply does not remove removable total 
derivative terms from the action. In Boulware et-al's \cite{b:a} 
prescription one has to quantize with basic variables. The basic 
variables are $\alpha, \alpha'$, if auxilliary variable is introduced in 
the vacuum Einstein-Hilbert action, while truely the basic variable in this 
situation is $\alpha$. Effectively, introduction of auxilliary variable 
in such situations, where it is not required, means to entertain 
redundant degree of freedom. So it was concluded that Boulware 
et-al's \cite{b:a} prescription should be taken up only after removing 
removable total derivative terms from the action. Unless this is done, wrong 
qusantum dynamics would emerge. In this sense all the auxilliary variables 
identified by Hawking and Luttrell \cite{h:n}, Horowitz \cite{h:p} and 
Pollock \cite{p:n} are wrong in the quantum domain.
\par
In a nutshell, in order to uniquely identify the auxilliary variable and 
to obtain the correct quantum description, we emphasize on the fact 
that Boulware et-al's \cite{b:a} prescription should be taken up only 
after removing the total derivative terms from the action. If this is 
done one can never get such variables to introduce in situations like 
vacuum Einstein-Hilbert action or induced theory of gravity and to end 
up with wrong quantum equations. On the other hand such variables can 
be introduced in the higher order gravity theory uniquely. Further, 
this method of identifying the auxilliary variable has got extremely 
interesting consequences as already mentioned at the beginning. It gives 
birth to a Schr$\ddot{o}dinger$-like equation where one of the variables 
being distangled from the kinetic part of the canonical variables act 
as the time parameter. Further an effective Hamiltonian emerges which is 
hermitian and as a 
result quantum mechanical idea of current and probability densities is 
realized. In addition the effective Hamiltonan contains an effective 
potential whose extremization yields classical constraint equation 
\cite{a:p} and \cite{a:c}. Horowitz \cite {h:p} claimed that the action 
for the higher order gravity theory does not require to have surface 
terms. This as a result of identifying the new variable properly, 
turns out to be wrong. This is due to the fact that removal of total 
derivative terms from the action leads to surface terms. One can get 
rid of these surface energy terms only if such counter terms are already 
included in the action. Tomboulis \cite{t:p} stated that higher order 
gravity theory has got the problem of unitarity at least in the 
perturbative models. We observed that \cite{a:p} and \cite{a:c}, when 
the classical Hamiltonian of $R+R^2$ gravity theory is quantized, an 
effective Hamiltonian emerges, which is hermitian at least in the 
Robertson-Walker minisuperspace model.      
\par
So far, there has been different strong motivations to modify 
Einstein-Hilbert action by higher order curvature invariant terms. To 
state a few, Stell 
\cite{k:p} has shown that the action $\int d^4{x}\sqrt{-g}[AC_{ijkl}+BR+CR^2 
]$is renormalizable in 4-dimensions. $C_{ijkl}$ is the Weyl tensor, $R$ 
is the Ricci scalar and $A,B,C$ are the coupling constants. Starobinsky 
\cite{st:p} obtained an inflationary solution without invoking phase 
transition in the very early Universe, from a field equation containing 
only geometric terms. Later, Starobinsky and Schmidt \cite{st:c} have shown 
that the inflationary phase is an attractor in the neighbourhood of the 
fourth order gravity theory. Hawking and Luttrell \cite{h:n} observed 
that Einstein-Hilbert action being modified by a curvature squared term 
is equivalent to the Einstein-Hilbert action being coupled to a massive 
scalar field. The Euclidean form of the Einstein-Hilbert action is 
not bounded from below, as a result the ground state wave function 
proposed by Hartle and Hawking \cite{hh:p} diverges badly. A remedy 
to this undesirable feature was suggested by Horowitz \cite{h:p}, who 
proposed a positive definite action in the form 
\be
S = \int~~d^4{x}\sqrt{g}[AC^2_{ijkl}+B(R-4\Lambda)^2]
\ee
$\Lambda$ being the cosmological constant. Not only that this action 
leads to a convergent functional intregral, it also reduces to the 
Einstein-Hilbert action in the weak field limit. A very recent and 
perhaps one of the most important motivation to modify Einstein-Hilbert 
action by the introduction of the curvature squared term is being put 
forward by Sanyal and Modak \cite{a:p}, \cite{a:c}. The motivation is to 
obtain a quantum mechanical interpretation of quantum cosmology. It is 
therefore required to extend our previous work by applying it to 
different situations to convincingly prove that our method is correct.
\par
In the present paper, we first take up the Induced Theory of Gravity in 
the following section and show that the corresponding Wheeler-deWitt 
equation is not satisfied by the Schr$\ddot{o}$dinger like equation 
obtained by Pollock \cite{p:n}, by introducing auxilliary variable.
This situation is considered to prove that auxilliary variable, if 
introduced in situations where it is not required, will yield wrong 
quantum dynamics. The reason behind is that, in the Boulware-et-al's 
\cite{b:a} proposal one has to finally quantize with basic variables. 
Introduction of auxilliary variables in such situations means to introduce 
redundant degrees of freedom. It further shows that once 
removable total derivative terms are removed from the action, 
auxilliary variables can not be introduced any more in such cases, 
and there is no possibility of emergence of wrong quantum 
description. 
\par
We then take up the modified Induced Theory of Gravity, 
by introducing $R^2$ term in the action, in section 3. This case was again 
studied by Pollock \cite{p:n}. We have shown that, following the method 
mentioned it is possible to distangle a combination of the 
expansion parameter and the scalar field from the kinetic part of 
the canonical variables that starts behaving as the time parameter 
of the corresponding quantum equation. Further, the effective 
Hamiltonian turns out to be hermitian and the quantum mechanical 
idea of current and probability densities follows naturally. 
These results are completely different from that obtained by 
Pollock \cite{p:n} and obcourse extremely illuminating.
\par
In section 4, The positive definite action proposed by Horowitz 
\cite{h:p} has been considered. In the Robertson-Walker 
minisuperspace model the Weyl tensor does not contribute. The 
quantum version of this model is found to tally with $R+R^2$ 
theory.
\par
In section 5, the next higher order curvature invariant term, viz., 
$R_{\mu\nu}R^{\mu\nu}$ is taken up to modify Einstein-Hilbert action. 
Interestingly enough, 
it has been found to yield the same type of classical and quantum 
results as was obtained in the $R^2$ case \cite{a:p}. It therefore 
appears that the hermiticity of the effective Hamiltonian operator 
and the quantum mechanical probabilistic interpretation 
perhaps are the generic feature of higher order gravity 
theory.
\par
Concuding remarks are made in section 6.               

\section{\bf{Induced Theory Of Gravity - A Toy Model}} As mentioned in the
introduction, Pollock \cite{p:n} introduced an auxilliary variable in the
induced theory of gravity to show that the corresponding Wheeler-deWitt
equation can well be written as Schr$\ddot{o}$dinger like equation. In
this section our aim is to show that not only it is unnecessary to
introduce auxilliary variable in such a theory but also it leads to wrong
quantum dynamics. For this purpose we shall deal with the same action and
write the corresponding Wheeler-deWitt equation in the standard procedure
without invoking an auxilliary variable, obtain the semi--classical
solution and compare it with that obtained by Pollock \cite{p:n}. Let us
start with the following action, \be
A=\int~~d^4{x}\sqrt{-g}[-\frac{1}{2}\epsilon \phi^2 R-\frac{1}{4}\lambda
\phi^4]. \ee In the Robertson-Walker mini-superspace model \be
ds^2=e^{2\alpha(\eta)}[-d\eta^2+d\chi^2+sin^2{\chi}(d\theta^2+sin^2{
\theta} d\phi^2)] \ee the above action takes the following form \be
A=-2\pi^2\int~~[3\epsilon\phi^2(1+\alpha'^2+\alpha'')e^{2\alpha}-\frac{1}{4}\lambda\phi^4
e^{4\alpha}]d\eta \ee where dash(') stands for derivative with respect to
the conformal time $\eta$. It further reads, \be A=2\pi^2\int~~[3\epsilon
(\phi^2(\alpha'^2-1)+2\phi\alpha'\phi')+\frac{\lambda}{4}\phi^4
e^{2\alpha})]e^{2\alpha}d\eta \ee apart from a total derivative term.
Corresponding classical field equaions are \be
\alpha''+\alpha'^2+2\alpha'\frac{\phi'}{\phi}+\frac{\phi''}{\phi}+\frac{\phi'^2}{\phi^2}-\frac{\lambda\phi^2
e^{2\alpha}}{6\epsilon}+1=0 \ee \be \alpha''+\alpha'^2-\frac{\lambda
\phi^2 e^{2\alpha}}{6\epsilon}+1=0 \ee \be
\alpha'^2+2\alpha'\frac{\phi'}{\phi}-\frac{\lambda\phi^2
e^{2\alpha}}{12\epsilon}+1=0 \ee In the phase space variable the
Hamiltonian is \be H=\frac{1}{36\epsilon^2\phi^2
e^{4\alpha}}[-p_{\phi}^2+\frac{2p_{\alpha}p_{\phi}}{\phi}-3\epsilon\phi^4
e^{6\alpha}+36\epsilon^2\phi^2 e^{4\alpha}]=0 \ee The corresponding
Wheeler-deWitt equation is \be
[\hbar^2(\frac{\partial^2}{\partial{\phi^2}}+\frac{q}{\phi}\frac{\partial}
{\partial{\phi}}-\frac{1}{\phi}\frac{\partial^2}{\partial{\alpha}\partial
{\phi}})+3\epsilon\phi^2e^{4\alpha}(12\epsilon-\phi^2 e^{2\alpha})]\psi=0 
\ee 
where $q$ is theoperator ordering index. The above Wheeler-deWitt equation 
admits the following semiclassical solution for $q = 0$.  
\be 
\psi=C~~exp[\frac{i}{\hbar}(n\phi^{l+1}+m) e^{l\alpha}] 
\ee 
where C, l, m, n are constants. 
\par 
Now we compare our result with that obtained 
by Pollock \cite{p:n} for the same action. Instead of removing surface term
from the action, Pollock introduced an auxilliary variable in the action
(4) as prescribed by Boulware et-al \cite{b:a}. Indeed the classical field
equations remain unchanged because the variables are canonical. The
Wheeler-deWitt equation obtained by Pollock is 
\be
i\frac{\partial\psi}{\partial\alpha}=\frac{1}{mx}\frac{\partial^2\psi}
{\partial{x^2}}+\frac{i}{x}\frac{\partial(1+x^2)\psi}{\partial{x}}+i(p-2)\psi.
\ee 
where $x=\alpha'$. This equation looks like Schr$\ddot{o}$dinger
equation which admits a semiclassical solution as was obtained by Pollock for
$p=2$, which is equivalent to our $q = 0$ situation, viz., 
\be 
\psi(\alpha,x)=exp[-im(x+\frac{1}{3}x^3)] \ee where
$m=\frac{72\pi^2 \epsilon^2}{\lambda}$. It is not difficult to see that
this Wheeler-deWitt equation is different from (10) with a different
solution altogether. So, the question is what went wrong ? This was
suggested in a couple of recent communications \cite{a:p}, \cite{a:c}. The
fact that all canonical variables lead to the same classical field
equations has no analogue in quantum domain. So to obtain the correct
quantum description one has to choose canonical variables carefully and
perhaps uniquely. This is achieved if such variabls are chosen only after
removing total derivative terms from the action. Unless otherwise  
 redundant degrees of freedom are taken into account, since in the 
Boulware-et-al's \cite{b:a} prescription one has to finally quantize with 
basic variables. We observe from action (5) that the basic variables  
are $\alpha$ and $ \phi$, while Pollock considered $\alpha$, 
$\alpha'$ and $\phi$ to be the basic variables which is definitely not 
true. As it is evident, that  removal of total derivative term 
from action (4) yields action (5) where there is no scope to introduce 
any auxilliary variable and as such unique quantum description emerges for 
a particular action.  
\section{\bf{Induced Theory Of Gravity Modified By Curvature Squared term}}
We now extend our work by introducing higher order term in the above action 
(2) along with a kinetic and potential term for the coupling scalar 
field. The action now reads, 
\be 
A=\int~~\sqrt{-g}[-\frac{1}{2}\epsilon\phi^2 R -
\frac{1}{2}\phi_{;\mu}\phi^{;\mu}-V(\phi) - \frac{\beta}{24}R^2]d^4 x 
\ee
which in the Robertson-Walker model takes the following form, \be
A=2\pi^2\int~~[(3\epsilon\phi^2
e^{2\alpha}-3\beta\alpha'^2-3\beta)\alpha''+3\epsilon\phi^2(\alpha'^2+1)e^{2\alpha}-\frac{3}{2}\beta(\alpha''^2+\alpha'^4+2\alpha'^2+1)+\frac{1}{2}\phi'^2
e^{2\alpha}-V(\phi)e^{4\alpha}]d\eta \ee The first three terms in the
action (15) are removable total derivative terms, which as pointed out
earlier leads to wrong Wheeler-deWitt equation when not removed. So our
first task is to integrate out these terms and then introduce auxilliary
variable and write the action in the canonical form as suggested by
Boulware et-al \cite{b:a}. So in the first step we obtain the following
action \be
A=2\pi^2\int~~[-3\epsilon\phi\alpha'(\phi\alpha'+2\phi')e^{2\alpha}-\frac{3}{2}\beta(\alpha''+\alpha'^4+2\alpha'^2+1)+3\epsilon\phi^2
e^{2\alpha}-V(\phi)e^{4\alpha}]d\eta + \Sigma_{1} \ee 
where $\Sigma_{1} = 2\pi^2[3\epsilon\phi^2\alpha' 
e^{2\alpha}-\beta\alpha'^3-3\beta\alpha']$ is the
surface term. In the next step we introduce the auxilliary variable, which
is \be Q=-\frac{1}{2\pi^2}\frac{\partial{A}}{\partial{\alpha''}} =
3\beta\alpha''. \ee In the following step we write the action (15) in the
canonical form where again one encounters another integrable term which
when integrated one obtains the following action. \be
A=2\pi^2\int~~[\alpha'Q'-\frac{3}{2}\beta\alpha'^2-3(\epsilon\phi^2
e^{2\alpha}+\beta)\alpha'^2-6\epsilon\phi\alpha'\phi'e^{2\alpha}+\frac{1}{2}\phi'^2
e^{2\alpha}+\frac{Q^2}{6\beta}+3\epsilon\phi^2
e^{2\alpha}-\frac{3}{2}\beta-V e^{4\alpha}] d\eta +\Sigma_{2} \ee where
$\Sigma_{2} = \Sigma_{1}-2\pi^2 \alpha' Q$ is the modified surface term. One 
can now write the classical 
field equations and verify after substituting the definition of $Q$ from
equation (17) that the equations are the same even if one would have
introduced the auxilliary variable in action (15) where total derivative
terms are not removed. However, since we are only interested in the
quantum dynamics for such an action, therefore we are writing only the
Hamiltonian constraint equation, which is \be
H=p_{\alpha}p_{Q}+\frac{p_{\phi}^2}{e^{2\alpha}}+6\epsilon\phi
p_{\phi}p_{Q}+\frac{3}{2}\beta p_{Q}^4 +3(\beta+\epsilon\phi^2
e^{2\alpha}+6\epsilon^2 \phi^2
e^{2\alpha})p_{Q}^2-\frac{Q^2}{6\beta}-3\epsilon\phi^2
e^{2\alpha}+\frac{3}{2}\beta+V(\phi)e^{4\alpha}. 
\ee 
In the above, $p_{\alpha} = \frac{\partial{L}}{\partial{\alpha'}}$, 
$p_{Q} = \frac{\partial{L}}{\partial{Q'}} = \alpha'$ and $p_{\phi} = 
\frac{\partial{L}}{\partial{\phi'}}$. Now as suggested by Boulware 
et-al \cite{b:a}, we quantize with basic variables, which are $\alpha, 
\alpha'$ and $\phi$. To avoid confusion we choose $\alpha' = x$ and 
replace $p_{Q} = \alpha'$ by $x$ and $Q$ by $-p_{x}$. The Hamiltonian 
thus reads,
\be
H=xp_{\alpha}+\frac{p_{\phi}^2}{2e^{2\alpha}}+6\epsilon\phi 
xp_{\phi}-\frac{p_{x}^2}{6\beta}+\frac{3}{2}\beta 
x^4+3[\beta+\epsilon\phi^2(1+6\epsilon)e^{2\alpha}]x^2-3\epsilon\phi^2 
e^{2\alpha}+\frac{3}{2}\beta+V(\phi)e^{4\alpha}.
\ee
This Hamiltonian which is constrained to vanish is now ready for 
quantization. The corresponding Wheeler-deWitt equation is
\be
i\hbar(\frac{\partial}{\partial{\alpha}}+6\epsilon\phi\frac{\partial}{\partial
{\phi}})\psi=\frac{\hbar^2}{6\beta}(\frac{1}{x}\frac{\partial^2}{\partial
{x}^2}+\frac{n}{x^2}\frac{\partial}{\partial{x}})\psi-\frac{\hbar^2}{2xe^
{2\alpha}}(\frac{\partial^2}{\partial{\phi^2}}+\frac{p}{\phi}\frac{\partial}
{\partial{\phi}})\psi+V_{e}\psi,
\ee
where, $n$ and $p$ are operator ordering indices. The effective potential 
$V_{e}$ is given by \be
V_{e}=\frac{3\beta}{2x}(x^2+1)^2+\frac{3\epsilon\phi^2 
e^{2\alpha}}{x}(2\epsilon x^2+x^2-1)+\frac{V(\phi)e^{4\alpha}}{x},
\ee  
Choosing the time variable as
\be
\frac{\partial}{\partial{t}}=\frac{\partial}{\partial{\alpha}}+6\epsilon\phi\frac{\partial}{\partial{\phi}}
\ee
the above Wheeler-deWitt equation can clearly be expressed as 
\be
i\hbar\frac{\partial{\psi}}{\partial{t}}=\hat{H}_{0}\psi.
\ee
This equation looks very similar to the Schr$\ddot{o}$dinger equation, 
where the Hamiltonian operator $\hat{H_{0}}$, given by the right hand 
side of equation (20) is clearly hermitian. This implies that we are now 
dealing with observables for which a probability interpretation exits. 
For $n=-1$ and $p=0$ the continuity equation is expressed as 
\be
\frac{\partial{\rho}}{\partial{t}}+\bf{\nabla}\cdot{\bf{J}}=0
\ee
where, $\rho = \psi^* \psi$, is the probability density and ${\bf{J}} = 
(j_{x},j_{\phi},0)$, is the current density, with $j_{x} = 
\frac{i\hbar}{6\beta x}(\psi^* \psi,_{x} - \psi\psi^*,_{x})$ and $j_{\phi} 
= -\frac{i\hbar e^{-2\alpha}}{2x}(\psi^* \psi,_{\phi} - 
\psi\psi^*,_{\phi})$. It is possible to write the continuity equation 
for other operator ordering indices also but that with respect to a 
different variable which is only functionally related to $x$. It is to 
be noted that for $\epsilon=0$ one can obtain the results 
corresponding to an action containing $R^2$ term minimally coupled  
with a scalar field \cite{a:p}. The form of the continuity equation (25) 
convincingly suggests that the choice of the time parameter in equation 
(23) is correct. Thus time parameter arises naturally in the quantum 
dynamics of the alternative theory of gravitation containing curvature 
squared term.   
\par
One can now extremize the effective potential, which is equivalent to 
extremize the effective action when Kinetic energy is neglected. the 
effective potential given by equation (22) is assymetric with respect to 
the expansion parameter $x=\alpha'$, ie. $V_{e} = -V_{e}$. Further it is 
unsatble at the short and long range limits of $x=\alpha'$. 
However, the classical cosmological models depict that the initial 
epoch of the evolution of the Universe, where quantum physics plays the 
dominating role, starts with a large expansion rate $x$. It slows down as 
Universe evolves and the infinity of the effective potential at small $x$ 
has nothing to do with the physical world since Universe by that time has 
entered the classical regime and the effective potential has no role at 
that stage.
\par
Extremization of the effective potential (22) with respect to $x$ yields, 
keeping $\alpha$ and $\phi$ fixed
\be
\beta(3x^4+2x^2-1)+2\epsilon\phi^2(2\epsilon 
x^2+x^2+1)e^{2\alpha}-\frac{2}{3}V(\phi)e^{4\alpha}=0
\ee
In the absence of the coupling parameter $\epsilon$ ie. the situation 
where the action is given by the curvature squared term minimally coupled 
with a scalar field, the above equation reads
\be
\beta(3x^4+2x^2-1)=\frac{2}{3}V(\phi)e^{4\alpha}
\ee
This is the classical constraint equation at a regime much below the 
Planck scale, where the Hamiltonian is mostly dominated by the potential 
energy, kinetic energy being much small.
\par
Further in the absence of the scalar potential $V(\phi)$ the equation 
(27) becomes 
\be
3x^4+2x^2-1=0
\ee
ie. either
\be
x^2+1=0
\ee
which is the vacuum Einstein's equation admitting Euclidean wormhole 
solution, or,
\be
3x^2-1=0
\ee
which implies that the extremum admits a solution 
$a=\frac{t-t_0}{\sqrt{3}}$, $a$ being the scale factor of the Universe in 
proper time $t$. For this solution the horizon radius $r_{H}$ is 
proportional to $ln(t)$. So at $t \rightarrow 0$, $r_{H} \rightarrow \infty$ 
and 
thus the horizon problem is solved. These results are obtained previously in 
\cite{a:p} and \cite{a:c}. It should be noted that $\beta$ can not be made 
zero at any stage after equation (16). This removes any possibilities of 
including redundant degrees of freedom, as  already mentioned. 
\par
Now extremum of the effective potential $V_{e}$ with respect to $\phi$ 
gives, keeping $\alpha$ and $x$ fixed
\be
6\epsilon\phi(2\epsilon x^2 +x^2 -1)+\frac{dV(\phi)}{d\phi}e^{2\alpha}=0
\ee
If the scalar field potential $V(\phi)$ has got an extremum at $\phi=0$ 
then the above equation is trivially satisfied, for which 
$\frac{\partial{V_{e}}}{\partial{x}}=0$ gives 
\be
\beta(x^2+1)(3x^2-1)=\frac{2}{3}e^{4\alpha}V(\phi)|_{extremum}
\ee
and
\be
\frac{\partial^2{V_{e}}}{\partial{x^2}}=6\beta(3x^2+1)
\ee
which is positive definite, implying extremum of $V_{e}$ with respect to 
$x$ has a minimum. Further,
\be
\frac{\partial^2{V_{e}}}{\partial{\phi^2}}=\frac{6\epsilon 
e^{2\alpha}}{x}(2\epsilon x^2+x^2+1)+\frac{e^{4\alpha}}{x}\frac{d^2 
V(\phi)}{d{\phi^2}}|_{extremum}
\ee
and finally,
\be
\frac{\partial^2{V_{e}}}{\partial{x}\partial{\phi}}=0
\ee
at the extremum. The condition that $V_{e}$ has an extremum being a 
function of two variables $x$ and $\phi$ is
\be
\frac{\partial^2{V_{e}}}{\partial{x^2}}\frac{\partial^2{V_{e}}}
{\partial\phi^2}-\frac{\partial^2{V_{e}}}{\partial{x}\partial{\phi}}>0
\ee
The left hand side turns out to be
\be
6\beta\frac{(3x^2+1)}{x}[6\epsilon(2\epsilon 
x^2+x^2+1)+e^{2\alpha}\frac{d^2{V(\phi)}}{d\phi^2}|_{extremum}]e^{2\alpha}
\ee
which is positive definite if the scalar field potential $V(\phi)$ has 
got a minimum, since $\epsilon > 0$, as long as $x > 0$ and real. It 
should be noted that $\alpha$ has not been considered as a variable, 
since as $\epsilon\rightarrow 0$, as well as when $\phi \rightarrow 0$, 
$\alpha$ alone 
acts as the time parameter. As the condition for the extremum has been 
fulfilled, if we now consider that the minimum of the scalar field 
potential $V(\phi = 0) = 0$, then equation (32) reduces to either 
\be
x^2+1=0
\ee
or
\be
3x^2-1=0
\ee
The same conditions that we have already visited. However, the first 
condition implies that $x$ is imaginary for which the condition for 
extremum of $V_{e}$ given in equation (37) is not satisfied. Hence, second 
condition  viz. equation (39) should 
only be considered keeping $x > 0$, ie. 
\be
x=\frac{1}{\sqrt{3}}
\ee
whose solution has already been given. We can thus conclude that the 
Universe is sitting at the extremum of the effective potential 
$V_{e}$, at the location of the configuration space variables $\phi = 
0$ and $x = \frac{1}{\sqrt{3}} $, which satisfies a solution $a = 
\frac{t-t_{0}}{\sqrt{3}}$, that solves the horizon problem, as 
already noticed.        

\section{\bf{Quantum Cosmology Of A Positive Definite Action}}
As already mentioned in the introduction, it is well known that the 
Euclidean form of the Gravitational Einstein-Hilbert action is 
not bounded from below. As a result the ground state wave funcction 
$\psi_{0}[h_{ij}] = N\int~~\delta{g} exp(-I_{E}[g])$ proposed by 
Hartle-Hawking \cite{hh:p} diverges. A remedy is to suggest a 
positive definite Euclidean action. Horowitz \cite{h:p} suggested 
an action (1) whose pseudo-Riemannian form is,
\be
A= \int~~\sqrt{-g}d^4 x[AC_{ijkl}+B(R-4\lambda)^2].
\ee
For the Robertson-Walker metric $C_{ijkl}$, the Weyl tensor vanishes   
and hence the above action can be expressed in the form,
\be
A=-\frac{\beta}{24}\int~~[R-4\lambda]^2\sqrt{-g}d^4 x.
\ee 
Our interest is to study the quantum dynamics of such a theory. In 
the Robertson-Walker minisuperspace model this action takes the 
following form 
\be
A=-\frac{m}{4}\int~~[\alpha''^2+(1+\alpha'^2)^2+\frac{4\lambda}{3}(1-\alpha'^2)e^{2\alpha}+\frac{4}{9}\lambda^2 e^{4\alpha}]~d\eta
\ee 
apart from a surface term 
$\Sigma_{1} = -\frac{m}{12}[6\alpha'+2\alpha'^3-4\lambda\alpha' 
e^{2\alpha}]$, where $m=12\beta\pi^2$. Now we define the auxilliary 
variable
\be
Q=-\frac{4}{m}\frac{\delta A}{\delta\alpha''}= ~~2\alpha''
\ee
and write the action in the canonical form after introducing the 
auxilliary variable as,
\be
A=\frac{m}{4}\int~~[\alpha'Q'-(\alpha'^2+1)^2+\frac{4\lambda}{3}(\alpha'^2-1)e^{2\alpha}+\frac{Q^2}{4}-\frac{4}{9}\lambda^2 e^{4\alpha}]~d\eta
\ee
apart from a modified surface term $\Sigma_{2} = -\frac{m}{6}
[3\alpha'+\alpha'^3-2\lambda\alpha'e^{2\alpha}+3\alpha'\alpha'']$. The 
Hamiltonian in the phase space variable can be written as
\be
P_{\alpha}P_{Q}+\frac{16}{m^2}P_{Q}^4+2(1-\frac{2\lambda}{3}e^{2\alpha})P_{Q^2}+m^2[\frac{\lambda^2}{36}e^{4\alpha}+\frac{\lambda}{12}e^{2\alpha}-\frac{1}{64}Q^2+\frac{1}{16}]=0
\ee
We now quantize the theory with basic variables $\alpha$ and $\alpha' = x 
(say)$. Therefore we replace as before, $Q$ by $-\frac{4}{m}P_{x}$ and 
$P_{Q}$ by $\frac{m}{4}x$. With such replacement the Hamiltonian constraint 
equation reads
\be
xP_{\alpha}=\frac{1}{m}P_{x}^2-\frac{m}{4}(x^2+1)^2+\frac{m\lambda}{3}(x^2-1)e^{2\alpha}-\frac{m}{9}\lambda^2 e^{4\alpha}.
\ee   
The corresponding Wheeler-deWitt equation is thus,
\be
i\hbar\frac{\partial\psi}{\partial\alpha}=\frac{\hbar^2}{mx}\frac{\partial^2\psi}{\partial x^2}+\frac{\hbar^2 n}{mx^2}\frac{\partial\psi}{\partial x}+[\frac{m}{4x}(x^2+1)^2-\frac{m\lambda}{3x}(x^2-1)e^{2\alpha}+\frac{m\lambda^2}{9x}e^{4\alpha}]\psi
\ee
where $n$ is the operator ordering index. As before the Wheeler-deWitt 
equation can be written as 
\be
i\hbar\frac{\partial\psi}{\partial\alpha}=\hat{H}_{0}\psi,
\ee
where, $\hat{H}_{0}\psi$ is the right hand side of the Wheeler-de-Witt 
equation (48). The probability interpretation follows naturally from the 
continuity equation, which can be written for $n = -1$ as,
\be
\frac{\partial\rho}{\partial\alpha}+\bf{\nabla}\cdot\bf{J}=0,
\ee
where  $\rho = \psi^* \psi$ is the probability density and ${\bf{J}} = 
(j_{x},0,0)$ is the current density, with 
$j_{x} = \frac{i\hbar}{mx}(\psi^*\psi_{,x} - \psi\psi^*_{,x})$. The 
continuity equation (50) proves that there is nothing wrong in 
considering $\alpha$ as the time parameter, and it appears naturally in 
the quantum dynamics of alternative theory of gravitational action 
containing curvature squared term. 
\par 
The effective potential in this case 
is \be
V_{e}=\frac{m}{4x}[(x^2+1)^2-\frac{4\lambda}{3}(x^2-1)e^{2\alpha}+\frac{4}{9}\lambda^2 e^{4\alpha}]
\ee
The extremum of the effective potential $V_{e}$ with respect to $x$ yields
\be
(x^2+1)(3x^2-1-\frac{4\lambda}{3}e^{2\alpha})-\frac{4}{9}\lambda^2 
e^{4\alpha}=o 
\ee
As it is evident, this is essentially the classical constraint equation 
at the epoch when the Universe is dominated by the potential energy term, 
kinetic energy being neglected. For vanishing $\lambda$ the extremum is 
either at
\be
x^2+1=0
\ee
which has the classical wormhole solution, or at
\be
3x^2-1=0
\ee
whose solution is $\alpha = ln(\frac{t-t_{0}}{\sqrt{3}})$. The 
consequences of these solutions are already discussed. All these results 
for $\lambda = 0$ are already obtained by Sanyal and Modak \cite{a:p}.
\section{\bf{Einstein-Hilbert Action Modified By $R_{\mu\nu}R^{\mu\nu}$ 
Term}}
In this section we couple the Einstein-Hilbert action with the next 
higher order curvature invariant term, viz. $R_{\mu\nu}R^{\mu\nu}$. Thus the 
action becomes \be
A=\int~~-\frac{\sqrt{-g}}{16\pi 
G}[R+\frac{\gamma}{2}R_{\mu\nu}R^{\mu\nu}]d^4 x
\ee
where $\gamma$ is the coupling constant, the Ricci scalar 
$R = - 6(\alpha'' + \alpha'^2 + 1)$  and $R_{\mu\nu} R^{\mu\nu} = 3 
(3\alpha''^2 + \alpha'' + 2\alpha'^2 + 2) e^{-4\alpha}$. Introducing  
$R$ and $R_{\mu\nu}R^{\mu\nu}$ and removing removable total derivative 
terms from the action, as proposed, we arrive at 
\be
A=M\int~~[-\gamma\alpha''^2-(\gamma(\alpha'^2+1)^2+(\alpha'^2-1)e^{2\alpha})]
d\eta+\Sigma_{1}
\ee
where $M=\frac{3\pi}{4G}$ and $\Sigma_{1} = \frac{3\pi}{4G} (\alpha' 
e^{2\alpha}- \gamma\alpha'- \frac{\gamma}{3}\alpha'^3 )$ is the 
surface term.
\par
Now introducing the new variable
\be
Q=\frac{1}{M}\frac{\partial{A}}{\partial{\alpha''}}=-2\gamma \alpha'
\ee
we express the action (56) in the following canonical form
\be
A=M\int~~[-\alpha'Q'+(1-\alpha'^2)e^{2\alpha}-\gamma(1+\alpha')^2
+\frac{Q^2}{4\gamma}]+\Sigma_{2}
\ee
Where $\Sigma_{2} = \Sigma_{1} + M\alpha' Q$ is the modified surface term. 
The classical field equations are
\be
\gamma(\alpha''''-6\alpha''\alpha'^2-2\alpha'')-(\alpha''+\alpha'^2+1)
e^{2\alpha}=0   
\ee 
and
\be
\gamma[2\alpha'''\alpha'-\alpha''^2-(1+\alpha'^2)(3\alpha'^2-1)]-(1+\alpha'^2)
e^{2\alpha}
\ee
The last equation is essentially the 
Hamiltonian constraint equation. It can easily be seen that both the 
equations reduce to the vacuum Einstein equations for the coupling 
constant $\gamma = 0$. Further, most interesting is the fact that the 
classical field equations for $R + R_{\mu\nu} R^{\mu\nu}$ action 
turns out the same as that for $R + R^2$ action, only having the 
difference in the coupling constant.   
\par
To quantize the system we express the Hamiltonian in the phase space variables
as
\be
H=\frac{1}{M}[-p_{\alpha}p_{Q}-\frac{M^2}{4\gamma}Q^2
+\frac{\gamma}{M^2}p_{Q}^4+2\gamma p_{Q}^2+M^2\gamma 
+(p_{Q}^2-M^2 )e^{2\alpha}]=0
\ee
since the Hamiltonan is constrained to vanish. As already mentioned we 
quantize with the basic variables $\alpha$ and $\alpha' = x$ (say). Thus 
the Hamiltonian Constraint equation takes the following look in terms of 
the basic variables
\be
-xp_{\alpha}=-\frac{1}{4M\gamma}p_{x}^2+\gamma 
M(x^2+1)^2+M(x^2-1)e^{2\alpha}=0
\ee
Hence the quantum dynamics of the system is determined by the following 
equation
\be
i\hbar\frac{\partial{\psi}}{\partial{\alpha}}=\frac{\hbar^2}{4M\gamma 
x}(\frac{\partial^2{\psi}}{\partial{x}^2}
+\frac{n}{x}\frac{\partial{\psi}}{\partial{x}})+V_{e}.
\ee
Where the effective action
\be
V_{e}=\frac{M}{x}[\gamma(x^2+1)^2+(x^2-1)e^{2\alpha}].
\ee 
Now equation (63) can be written as 
\be
i\hbar\frac{\partial{\psi}}{\partial{\alpha}}=\hat{H}_{0}\psi
\ee
Where, $\hat{H}_{0}$, the effective Hamiltonian operator operating on 
$\psi$, being given by the right hand side of equation (63), is again found 
to be hermitian. As a result one can easily construct the continuity 
equation
\be
\frac{\partial{\rho}}{\partial{\alpha}}+\bf{\nabla}\cdot\bf{J}
\ee
where, the current density $\bf{J} = (J_{x},0,0)$, $J_{x} = 
\frac{i\hbar}{4M\gamma 
x}(\psi^*\psi,_{x} - \psi\psi^*,_{x})$ and the probability density$\rho = 
\psi^*\psi$. Though 
the above continuity equation is formed for $n = -1$, yet it is not 
difficult to see that it is true in general for arbitrary operator 
orering index but then with respect to a new variable which is 
functionally related to $x$. In anlogy to quantum mechanics it is now quite 
trivial to identify $\alpha$ as the time parameter in quantm cosmology 
while the variable $x = \alpha'$ acts as the spatial coordinate, $\alpha$ 
being the  expansion parameter. It appeares  that the effective 
Hamiltonian operator $\hat{H}_{0}$ diverges at the bounce of the 
Universe, ie. at $\alpha' = 0$. It can be interpreted as the Universe 
enters into the Classical regime much before $\alpha' = 0$ and so 
effective Hamiltonian has no role to play at that regime. It can also be 
interpreted as the arrow of time reverses it's direction at the bounce.
\par
In the weak energy limit, the contribution of kinetic energy is pretty 
small with respect to the potential energy in the Hamiltonian $H_{0}$. 
At that regime the Hamiltonian is almost dominated by the potential 
energy, and so one can extremize the effective potential for $k = 1$ to get 
either 
\be
x^2+1=0
\ee
which is the vacuum Einstein's equation admitting Euclidean wormhole 
solution as already discussed, or
\be
\gamma(3x^2-1)+e^{2\alpha} = 0 
\ee 
which has a solution
\be
e^{\alpha} = a = \sqrt{\gamma} sin(\frac{t-t_{0}}{\sqrt{3\gamma}})
\ee
where $a$ is the scale factor in proper time $t$. It is interesting to note 
that all these results are at par with those obtained by Sanyal and Modak 
\cite{a:c} for the  Einstein-Hilbert action modified by the curvature squared 
term ignoring the scalar field. Altogether, we observe that in the 
Robertson-Walker minisuperspace model the Einstein-Hilbert action 
modified by the next higher order curvature invariant term 
$R_{\mu\nu} R^{\mu\nu}$  contributes nothing more than that modified by 
the $R^2$ term both at the classical and the quantum regime. However, 
it has been interestingly observed that the inclusion of such higher order 
terms in the Einstein-Hilbert action naturally gives rise to a 
Schr$\ddot{o}$dinger like equation with an effctive Hamiltonian which is 
hermitian. As a result, quantum mechanical probability interpretation of 
quantum cosmology is available.  
\section{\bf{Concluding Remarks}}
The aim of this paper is to show how Gravitational action containing 
higher order terms or being modified by such terms leads to a quantum 
mechanical probability interpretation of quantum cosmology. In a couple 
of recent publications \cite{a:p} and \cite{a:c} it has been shown that 
to quantize Gravitational action containing curvature squared term or 
being modified by such term, Boulware et-al's \cite{b:a} prescription can 
be taken up only after removing removable total derivative terms from the 
action. In \cite{a:p} a toy model viz. Einstein-Hilbert action was 
considered to show that if removable total derivative terms are kept 
in the action, one can introduce auxilliary variable , that leads to 
wrong quantum dynamics. Here again, it has been shown that it is possible 
to introduce auxilliary variable in the induced theory of gravity which 
ultimately leads to wrong quantum description. Here it has been pointed out 
that, introduction of auxilliary variables in situations where it is not  
required, effectively implies to consider redundant degrees of freedom 
during quantization. This is because, in Boulware et-al's \cite{b:a} 
prescription one finally has to quantize with basic variables. It has 
been shown in section 3 that, while induced theory of gravity has actually 
a pair of basic variables viz. $\alpha , \phi$, introduction of 
auxilliary variable demands basic variables to be three, viz. $\alpha, 
\phi, \alpha'$. Hence we conclude that all earlier works of choosing 
auxilliary variables arbitrarily or through some definite prescription 
lead to wrong quantum dynamics unless it has been introduced only 
after removing removable total derivative terms from the action.
\par 
We have shown that once removable total derivative terms are removed from 
the action, there is no chance of introducing such auxilliary variables 
in situations where not required. Further, such variables are thus chosen 
uniquely in situations where required, leading to unique quantum 
desccription of the theory.
\par
Such method of choosing auxilliary variables gives rise to a 
Schr$\ddot{o}$dinger like equation , where a time parameter comes out 
automatically. Further, it gives rise to an effective Hamiltonian which 
is hermitian and thus one can write the coontinuity equaation in view of 
the probability and current densities leading to a quantum mechanical 
probability interpretation of the theory. 
\par
The same type of result has been obtained by introducing the next higher 
order curvature invariant term $R_{\mu\nu}R^{\mu\nu}$ in the  
Einstein-Hilbertaction. This implies that quantum mechanical probability 
interpretation of quantum cosmology might be a generic feature of higher 
order curvature terms.
\par
It should be mentioned that all these results are valid only in the 
isotropic and homogeneous minisuperspace models. How to extend our work 
in the anoisotropic models is under investigation.

\end{document}